\documentclass[graybox]{svmult}

\usepackage[colorlinks=true,citecolor=green,linkcolor=red,filecolor=cyan,urlcolor=magenta]{hyperref}

\usepackage{epsfig}
\usepackage{color,amssymb}
\usepackage{amsfonts,amssymb}
\usepackage{amsmath,graphicx}
\usepackage{youngtab}
\usepackage[mathscr]{eucal}
\usepackage{multicol}
\usepackage{color}
\usepackage{graphicx}
\usepackage{pstcol}
\usepackage{pst-coil}
\usepackage{colortbl}
\usepackage[all]{xy}
\usepackage{epsfig}
\usepackage{color,amssymb}
\usepackage{amsfonts,amssymb}
\usepackage{amsmath}
\usepackage{youngtab}

\def\a{\alpha}
\def\b{\beta}

\def\g{\gamma}

\def\ve{\varepsilon}

\usepackage{mathptmx}       
\usepackage{helvet}         
\usepackage{courier}        
\usepackage{type1cm}        
\usepackage{makeidx}         
\usepackage{graphicx}        
\usepackage{multicol}        
\usepackage[bottom]{footmisc}

\makeindex             

\begin{document}

\title*{Axion and dilaton + metric emerge jointly from
  an electromagnetic model universe\\ with local and linear response
  behavior}

\titlerunning{Axion and dilaton + metric emerge from local and linear
  electrodynamics} 
\author{Friedrich W.\ Hehl}

\authorrunning{Axion and dilaton + metric emerge from local and linear
  electrodynamics}

\institute{Friedrich W.\ Hehl \at Univ.\ of Cologne and Univ.\ of
  Missouri-Columbia, email: {hehl@thp.uni-koeln.de}.  \at Invited
  contribution to T.~Asselmeyer-Maluga (ed.), {\it At the Frontier of
    Spacetime: Scalar-Tensor Theory, Bell's Inequality, Exotic
    Smoothness.}  Festschrift on the occasion of Carl Brans' 80th
  birthday, to be published by Springer in May 2016.}
%
%
\maketitle

\abstract{We take a quick look at the different possible universally
  coupled scalar fields in nature. Then, we discuss how the gauging of
  the group of scale transformations (dilations), together with the
  Poincar\'e group, leads to a Weyl-Cartan spacetime structure. There
  the {\it dilaton} field finds a natural surrounding. Moreover, we
  describe shortly the phenomenology of the hypothetical {\it axion}
  field. --- In the second part of our essay, we consider a spacetime,
  the structure of which is exclusively specified by the premetric
  Maxwell equations and a fourth rank electromagnetic response tensor
  density $\chi^{ijkl}= -\chi^{jikl}= -\chi^{ijlk}$ with 36
  independent components. This tensor density incorporates the
  permittivities, permeabilities, and the magneto-electric moduli of
  spacetime. No metric, no connection, no further property is
  prescribed. If we forbid birefringence (double-refraction) in this
  model of spacetime, we eventually end up with the fields of an
  axion, a dilaton, and the 10 components of a metric tensor with
  Lorentz signature. If the dilaton becomes a constant (the vacuum
  admittance) and the axion field vanishes, we recover the Riemannian
  spacetime of general relativity theory. Thus, the metric is
  encapsulated in $\chi^{ijkl}$, it can be {\it derived} from it.\\
  \underline{\hspace{0pt}} \hfill {\it file
    CarlBrans80$\underline{\hspace{3pt}}$08.tex, 26 Mar 2016} }

\newpage
\section*{Contents}
\begin{enumerate}
\item Dilaton and axion fields
\begin{enumerate}
\item[1.1] Scalar fields
\item[1.2] Einstein gravity and the energy-momentum current
\item[1.3] Einstein-Cartan gravity: the additional spin current
\item[1.4] Dilaton field and dilation current
\item[1.5] The Weyl-Cartan spacetime as a natural habitat of the dilaton field
\item[1.6] Axion field 
\end{enumerate}

\item An electromagnetic model universe
\begin{enumerate}
\item[2.1] The premetric Maxwell equations
\item[2.2] A local and linear electromagnetic response
\item[2.3] Propagation of electromagnetic disturbances
\item[2.4] Fresnel wave surface 
\item[2.5] Suppression of birefringence: the light cone
\item[2.6] Axion, dilaton, metric
\end{enumerate}
\item Discussion
\end{enumerate}
\underline{\hspace{0pt}}\vspace{30pt}

\hfill\begin{minipage}{90mm}
{\it ``Universally coupled, thus gravitational, scalar fields are still
active players in contemporary theoretical physics. So, what is the
relationship between the scalar of scalar-tensor theories, the dilaton
and the inflaton? Clearly this is an unanswered and important
question. The scalar field is still alive and active, if not always
well, in current gravity research.''}\medskip

\hfill Carl H.\ Brans (1997)

\end{minipage}

\section{Dilaton and axion fields}

\subsection{Scalar fields}

The Jordan-Brans\footnote{Carl Brans is one of the pioneers of the
  scalar-tensor theory of gravitation. This essay is dedicated to Carl
  on the occasion of his 80th birthday with all best wishes to him and
  his family. During the year of 1998, we had the privilege to host
  Carl, as an Alexander von Humboldt awardee, for several months at
  the University of Cologne. I remember with pleasure the many lively
  discussions we had on scalars, on structures of spacetime, on
  physics in general, and on various other topics.}-Dicke {\it
  scalar}, the {\it dilaton}, the {\it axion}, the {\it
  inflaton}---scalar fields everywhere---and eventually even one, the
scalar, that is, the spinless {\it Higgs} boson $ {H}^{ {0}}$, which
has been found experimentally as heavy as some 134 protons. These
different scalar fields\footnote{We skip here the plethora of {\it
    scalar} mesons,
  \begin{eqnarray*}&& {\pi^\pm,\pi^0,\eta,f_0(500), \eta'
      (958),f_0(980),a_0(980),...,} \\ && K^\pm,K^0,K^0_{S},
    K^0_{L},K^*_0(1430), {D^\pm,D^0,D^*_0 (2400)^0,D^\pm_s,...\,;}
\end{eqnarray*}
they are all composed of two quarks. Thus, the scalar mesons do not
belong to the fundamental particles.} are not necessarily independent
from each other, it could be, for example, that the JBD-scalar can be
identified with the dilaton (see \cite{Fujii:2003pa}) or the Higgs
boson with the inflaton (see
\cite{Barvinsky:2009fy,Bezrukov:2015}). Thus, the list of potentially
existing universally coupled scalar fields could be somewhat
smaller. For the history of the JBD-scalar, one can compare Brans
\cite{Brans:2005ra} and Goenner \cite{Goenner:2012cq} and, for the role
of the inflaton in different models, Vennin et
al.\ \cite{Vennin:2015eaa}.

\subsection{Einstein gravity and the energy-momentum current}

As remarked by Brans \cite{Brans:1996tp} in the quotation above, if
universally coupled, the scalar fields are intrinsically related to
the gravitational field. In Einstein's theory of gravity, general
relativity (GR), the gravitational potential is the metric $g_{ij}$,
with $i,j =0,1,2,3$ as (holonomic) coordinate indices. As its source
acts the symmetric energy-momentum tensor $T_{ij}$ of matter. This is
a second rank tensor, which is generated already in special relativity
(SR) with the help of the group $T(4)$ of {\it translations} in space
and in time.  Together with the Lorentz transformations $SO(1,3)$, the
translations $T(4)$ build up the Poincar\'e group $P(1,3)$ as a
semi-direct product: $P(1,3)=T(4)\rtimes SO(1,3)$. This is the group of
motion in the Minkowski space of special relativity, see
\cite{Giulini:2008}.

Accordingly, if one desires to understand gravity from the point of
view of the gauge principle, the $T(4)$ is an indispensable part of
these considerations. However, being only {\it one} piece of the
$P(1,3)$, it is suggestive to gauge the complete $P(1,3)$. This is
exactly what Sciama and Kibble did during the beginning 1960s, see
\cite{Obukhov:2006gea}, \cite[Chap.4]{Reader}, and
\cite{Chen:2015vya}.

\subsection{Einstein-Cartan gravity: the additional spin current}

This gauging of the $P(1,3)$ extends the geometrical framework of
gravity. The 4 translational potentials $e_i{}^ \alpha$ and the 6
Lorentz potentials $\Gamma_i{}^{\alpha\beta}=-\Gamma_i{}^{\beta\alpha}
$ span a Riemann-Cartan spacetime, enriching the Riemannian spacetime
of GR by the presence of Cartan's torsion; here $\alpha,\beta=0,1,2,3$
are (anholonomic) frame indices. Whereas the translational potentials
couple to the canonical energy-momentum tensor of matter
$\Sigma_\alpha{}^i$, the Lorentz potentials couple to the canonical
spin current of matter
$\tau_{\alpha\beta}{}^i=-\tau_{\beta\alpha}{}^i$.

The simplest version of the emerging Poincar\'e gauge models is the
Einstein-Cartan theory (EC), a viable gravitational theory competing
with GR if highest matter densities are involved. If
$l_{\text{Planck}}$ denotes the Planck length and
$\lambda_{\text{Compton}}$ the Compton wave length of a particle, then
deviations of the Einstein-Cartan from Einstein's theory are expected
at length scales of below
$\sim\!(l_{\text{Planck}}^2\,\lambda_{\text{Compton}})^{1/3}$; for
protons, prevalent in the early cosmos, it is about
$10^{-29}\,\text{meter}$. According to Mukhanov \cite{Mukhanov}, it is
exactly this order of magnitude down to which, according to recent
cosmological data, GR is known to be valid.

From a gauge theoretical point of view, the EC-theory looks more
reasonable than GR since the Einsteinian principles of how to
heuristically derive a gravitational theory were followed closely:
they were just applied to fermionic matter instead of to macroscopic
point particles or Euler fluids or to classical electromagnetism, as
Einstein did.

Incidentally, in the EC-theory and, more generally, in the Poincar\'e
gauge theory, the Poincar\'e and, in particular, the Lorentz
covariance are valid locally by construction, similar as in a $SU(2)$
Yang-Mills theory, we have local $SU(2)$ covariance. Kosteleck\'y
(priv.\ comm., Jan.\ 2016) agrees that the ``Einstein-Cartan theory
maintains local Lorentz invariance.'' Then the same is true for a
Poincar\'e gauge theory, which acts likewise in a Riemann-Cartan
spacetime with torsion and curvature. However, in an experimental
set-up, according to Kosteleck\'y \cite{Kostelecky:2007kx}, torsion
must be considered as an external field and, according to the
``standard lore for backgrounds,'' local Lorentz invariance is
broken. By the same token, an external magnetic field in
electrodynamics breaks local Lorentz invariance. This is, in my
opinion, an abuse of language, which conveys the wrong message that
the existence of a torsion field violates local Lorentz
invariance.

If, for theoretical reasons, one wants to evade the emergence of the
Riemann-Cartan spacetime, then one can manipulate, in the underlying
Minkowski space, the intrinsic or spin part of the total angular
momentum of matter in such a way that it vanishes on the cost of
increasing the orbital part of it by the corresponding amount, see
\cite{Mielke:1989vg}. This procedure is called Belinfante-Rosenfeld
symmetrization of the canonical energy-momentum current, which, in
general, is defined as an asymmetric tensor by the Noether
procedure. Accordingly, by symmetrization the energy-momentum current
is made fit to act as a source of the Einstein field equation. In this
way, one can effectively sweep the spin and the torsion under the rug
and can live happily forever in the paradise of the Riemannian
spacetime of GR.

Of course, in the end observations and/or experiments will decide
which of the two theories, GR or EC, will survive. We opt for the
latter.

\subsection{Dilaton field and dilation current}

The dilaton field $\phi$ entered life as a Nambu-Goldstone boson of
broken scale invariance, see Fujii in \cite{Fujii:2003pa}. Thus,
$\phi$ is related to dilat[at]ions or scale transformations in space
and time. But the dilaton also occurs in theories of gravity (JBD) and
in string theory, see Di Vecchia et al.\ \cite{DiVecchia:2015jaq}. The
$P(1,3)$, if multiplied (semi-directly) with the scale group, becomes
the Weyl group $W(1,3)$. This 11-parameter group is an invariance
group for massless particles in special relativity. The translations,
via Noether's theorem, generate the conserved energy-momentum tensor
$\Sigma_{ij}$, the Lorentz transformations the conserved total angular
momentum tensor
$J_{ij}{}^k:=\tau_{ijk}+x_{[i}\Sigma_{j]}{}^k=-J_{ji}{}^k$, and the
dilation the conserved total {dilation current}
$\Upsilon^k:=\Delta^k+x^l\Sigma_l{}^k$,
\begin{eqnarray}
\partial_k\Sigma_i{}^k&=&0\,,\\
\partial_kJ_{ij}{}^k&=&\partial_k\tau_{ij}{}^k-\Sigma_{[ij]}=0\,,\\
\partial_k\Upsilon^k&=&\hspace{5pt}\partial_k\Delta^k+\Sigma_k{}^k=0\,,
\end{eqnarray}
see \cite{Lopu,Kopczynski:1988jq} particularly for $\Upsilon^k$. Thus,
if a universal coupling is assumed, then $\phi$ should have the
intrinsic dilation current $\Delta^k$ as its source; for theories in
Weyl spaces in which $\Delta^k$ does {\it not} play a role, see Scholz
\cite{Scholz:2014kba,Scholz:2014tba}.

There are numerous field theoretical models under way which, if scale
or dilation invariance is implemented, have conformal invariance as a
consequence; for a more recent review see Nakayama
\cite{Nakayama:2015}. Hence, jumping to conformal invariance, before
one understood scale invariance, is probably not a very good
strategy. For this reason we confine ourselves here to scale
invariance, to the dilaton, and to the 11-parametric Weyl group. But
it should be understood that the light cone is also invariant under
the 15-parametric conformal group, see Barut \& R\c{a}czka
\cite{Barut} and Blagojevi\'c \cite{Blagojevic:2002} and, for a
historical account, Kastrup \cite{Kastrup:2008}.

Both currents, the intrinsic dilation current $\Delta^k$ and the
energy-momentum current $\Sigma_i{}^k$ are related to {\it external}
groups, to the dilation (scale) and to the translation groups,
respectively. This is the reason for their universality.

\subsection{The Weyl-Cartan spacetime as a natural habitat of the
  dilaton field}

We only tried to make a strong case in favor of the EC-theory in
order to repeat the corresponding arguments for the dilation
group. Gauging the Weyl group yields a Weyl-Cartan spacetime. The
classical paper in that respect is the one of Charap and Tait, see
\cite[Chap.8]{Reader}. A universally coupled massless scalar field
induces a {\it Weyl covector} $Q_i$ as the corresponding dilation
potential willy nilly. This is the type of spacetime Weyl used (with
vanishing torsion) for his failed unified theory of 1918. Here the
Weyl space with the connection $ ^{\text{W}}\Gamma$ is resurrected for
the dilation current, instead of for the electric current, see
\cite{Blagojevic:2002}:
\begin{equation}
  \stackrel{\rm W}{\nabla}_ig_{jk}=-Q_ig_{jk}\,,\quad\,
  ^{\text{W}}\Gamma_{ijk}=\,^{\text{RC}}\Gamma_{ijk}+\frac
  12(Q_ig_{jk}  +Q_jg_{ki}-Q_kg_{ij})\,;
\end{equation}
$^{\text{RC}}\Gamma$ is the connection of the Riemann-Cartan
space. Again, as in the case of the Lorentz group, one can manipulate
the total dilation current $\Upsilon^k$ and can transform its
intrinsic part into an orbital part by modifying in this case the {\it
  trace} $\Sigma_k{}^k$ of the energy-momentum current.  Then, again,
one can stay within the realm of the Riemannian space of GR, see
Callan, Coleman, and Jackiw \cite{Callan:1970ze}.

As we mentioned already, the gauge theoretical answer was given by
Charap and Tait \cite{Charap:1973fi}. Again, which approach will
succeed is eventually a question to experimental verification.

We see, if the JBD-scalar is interpreted as a dilaton, then we would
expect that the Weyl-Cartan spacetime is its arena. Clearly this does
only provide the kinematics of the theory. The dynamics would depend
on the exact choice of the dilaton Lagrangian. 

Recently, {Lasenby and Hobson} \cite{Lasenby:2015dba} wrote an
in-depth review of gauging the Weyl group and, moreover, formulated an
``extended Weyl gauge theory.'' Also within their framework, the
Weyl-Cartan space, and a straightforward extension of it, play an
important role, see also Haghani et al.\
\cite{Haghani:2014zra}. Definite progress has also been achieved in
the study of equations of motion within the scalar tensor theories of
gravity, see Obukhov and Puetzfeld
\cite{Obukhov:2014mka,Puetzfeld:2014qba,Puetzfeld:2015jha}. The
breaking of scale invariance in the more general approach of
metric-affine gravity was studied in \cite{Hehl:1989ij}, for example;
for somewhat analogous breaking mechanisms, see
\cite{Mielke:2011zza,Mielke:2011zz,Mielke:2013mla}.

\subsection{Axion field}

Dicke did not only introduce in 1961, together with Brans
\cite{Brans:1961sx}, a scalar field into gravity, but he also
discussed, in 1964, and pseudoscalar or axial scalar field $\varphi^2$
in the context of gravitational theory, see \cite[Appendix 4, p.51,
Eq.(7)]{Dicke:1964}.

Subsequently, in the {early} 1970s, Ni \cite{Ni:1973} investigated
matter coupled to the gravitational field and to electromagnetism and
looked for consistency with the equivalence principle. He found it
possible to introduce in this context a new neutral {\it
  pseudo\/}scalar field accompanying the metric field, see also
\cite{Ni:1977zz, Ni:1984,Balakin:2009rg,Ni:2014cca}. Later, in the
context of the vacuum structure of quantum chromodynamics, a light
neutral psedoscalar, subsequently dubbed ``axion'' was hypothesized,
see Weinberg \cite[pp.458--461]{Weinberg}. Similar as Ni's field, the
axion couples also to the electromagnetic field, see Wilczek's paper
\cite{Wilczek:1987mv} on ``axion electrodynamics''.

The axion field is of a similar universality as the gravitational
field. In other words, the axion belongs to the universally coupled
scalar fields. Let in electrodynamics, ${\cal
  H}^{ij}=(\vec{D},\vec{H})=-{\cal H}^{ji}$ and
$F_{ij}=(\vec{E},\vec{B})=-F_{ji}$ denote the excitation and the field
strength, respectively. The constitutive relation characterizing the
axion field $\alpha(x)$ (in elementary particle terminology it is
called $A^0$) reads \cite{birkbook},
\begin{equation}
  \mathcal{H}^{ij}=\frac
  12\alpha\epsilon^{ijkl}
F_{kl}\quad\text{or}\quad\begin{cases}
      D^a=\;\;\,\alpha B^a\,,\\ H_a=-\alpha E_a\,,
\end{cases}\label{axax}
\end{equation}
see also \cite{Hehl:2007ut} for the corresponding formalism; here
$\epsilon$ is the totally antisymmetric Levi-Civita symbol with
$\epsilon^{ijkl}=\pm 1$, moreover, $a=1,2,3$. Clearly, the axion
embodies the magnetoelectric effect par excellence. It is a pseudoscalar
under 4-dimensional diffeomorphisms.

In electrotechnical terms, the axion behaves like the (nonreciprocal
Tellegen) gyrator of network analysis, see \cite{Kupf,Russer}; also the
perfect electromagnetic conductor (PEMC) of Lindell \& Sihvola
\cite{PEMC,Sihvola:2008} represents an analogous
structure. Metaphorically speaking, as we see from (\ref{axax}), the
axion ``rotates'' the voltages $(\vec{B},\vec{E})$ into the currents
$(\vec{D},\vec{H})$. In SI, we have the units $[B]={\rm V}{s}
/{m^2}\,,[E]={\rm V}/m\,;\,\; [D]=As/{m^2}\,,[H]={\rm A}/m$. Thus,
$[\alpha]=1/\text{ohm}=1/\Omega$ carries the physical dimension of an
admittance. Now, in the Maxwell Lagrangian, we find an additional
piece $\sim\alpha(x) \epsilon^{ijkl}
F_{ij}F_{kl}\sim\alpha(x)\vec{E}\cdot\vec{B}$, a term, which was
perhaps first discussed by Schr\"odinger \cite[pp.25 to 26]{Schro}. If
$\alpha$ were a constant, the field equations would not change.

As we already remarked, $\alpha$ is a 4-dimensional pseudoscalar. The
same is true for the von Klitzing constant $R_{\text{K}}\approx
25\,813\,\Omega$. And this covariance is a prerequisite for its
universal meaning. Phenomenologically, the quantum Hall effect (QHE)
can also be described by a constitutive law of the type (\ref{axax}),
see \cite[Eq.(B.4.60)]{birkbook}. 

It is possible to apply the constitutive relation (\ref{axax})
directly to a solid, too. By the evaluation of experiments we have
shown \cite{Hehl:2007ut} that in the multiferroic Cr$_2$O$_3$
(chromium sesquioxide) we have a nonvanishing axion piece of up to
$\sim 10^{-3}\lambda_0$, where $\lambda_0$ is the vacuum admittance of
about $1/377 \,\Omega$. This fact demonstrates that there exist
materials with a nonvanishing, if small, (pseudoscalar) axion
piece. This may be considered as a plausibility argument in favor of a
similar structure emerging in fundamental physics. If the $A^0$ were
found, it would {\it not} be an unprecedented structure, see in this
context also Ni et al.\ \cite{Ni:1999di}.

In matter-coupled ${\cal N}=2$ supergravity models, there are examples
in which a dilaton and an axion are contained simultaneously in the
allowed particle spectrum, see Freeman and Van Proeyen
\cite[p.451]{Freedman}. However, in the next section we will
demonstrate that in a fairly simple classical model of an
electromagnetic universe, the axion can emerge jointly with a dilaton
and the metric.

More recently, there have been attempts to relate the axion field to
the torsion of spacetime, see, for example Mielke et al.\
\cite{Mielke:2006zp} and Castillo-Felisola et al.\
\cite{Castillo:2015ema}. To us, this assumed link between the internal
symmetry $U(1)$ of the axion with the external translation symmetry
$T(4)$ related to the torsion appears to be artificial and not
supported by physical arguments.

\section{An electromagnetic model universe}

\subsection{The premetric Maxwell equations}

We consider a 4-dimensional differentiable manifold. The
electromagnetic field is specified by its excitation
$\mathcal{H}^{ij}$, a 2nd rank antisymmetric contravariant tensor
density, and by its field strength $F_{ij}$, a 2nd rank antisymmetric
covariant tensor; the electric current $\mathcal{J}^k$ is a
contravariant vector density, see Post \cite{Post}. On this manifold,
the Maxwell equations read
\begin{equation}\label{Maxwell1}
  \partial_k\mathcal{H}^{ik}=\mathcal{J}^k\,,\qquad \partial_{[i}
  F_{jk]}=0\,;
\end{equation}
the brackets $_{[\;]}$ denote antisymmetrization of the corresponding
indices with ${1}/{3!}$ as a factor, see \cite{Schouten}; for the
Tonti-diagram of (\ref{Maxwell1}), compare \cite[p.315]{Tonti}.

In none of these equations the metric tensor $g_{ij}$ nor the
connection $\Gamma_{ij}{}^k$ are involved. Still, these equations are
valid and are generally covariant in the Minkowski space of special
relativity, in the Riemann space of general relativity, and in the
Riemann-Cartan or Weyl-Cartan space of gravitational gauge
theories. The Maxwell equations (\ref{Maxwell1}) as such, apart from a
historical episode up to 1916, see \cite{Einstein:1916Max,Meaning},
have no specific relation to the Poincar\'e or the Lorentz group.

Perlick \cite{Perlick:2010ya} has shown that the initial value problem
in electrodynamics can be particularly conveniently implemented by
means of the premetric form of the Maxwell equations.

In contrast to most textbook representations, no ``comma goes to
semicolon rule'' is required. The Maxwell equations (\ref{Maxwell1})
are just universally valid for all forms of electrically charged
matter. Incidentally, this represents also a simplifying feature for
numerical implementations. The price one has to pay is to introduce,
as Maxwell did, the excitation $\mathcal{H}^{ij}$, besides $F_{ij}$,
as an independent field quantity and to note that it is a tensor
density. From a phenomenological point of view, this is desirable
anyway, since the excitation has an operational definition of its own,
namely as charge/length$^2$ ($\vec{D}$) and current/length
($\vec{H}$), respectively, which is independent from the definition of
the field strength as force/charge ($\vec{E}$) and force/current
($\vec{B}$). For a rendition in the calculus of exterior
differentiable forms, one can compare with the axiomatic scheme in
\cite{birkbook} and \cite{Delphenich:2009}, see also
\cite{Delphenich:2015}.

Let us stress additionally that $\mathcal{H}^{ij}$, $F_{ij}$, and
$\mathcal{J}^k$ can be defined in a background independent way.

The Maxwell equations (\ref{Maxwell1}) are based on the conservation
laws of electric charge $Q:=\int
d\sigma^{ijk}\varepsilon_{ijkl}\,\mathcal{J}^l$ (unit in SI ``coulomb'')
and magnetic flux $\Phi:=\int d\sigma^{ij}F_{ij}$ (unit in SI
``weber''). Charge $Q$ and flux $\Phi$ are 4-dimensional scalars. They
induce the structure of the excitation $\mathcal{H}_{ij}$ and the
field strength $F_{ij}$. In this context, the field strength is
operationally defined via the Lorentz force density
$\mathfrak{f}_i=F_{ij}\mathcal{J}^j$, the current being directly
observable and the force and its measurement known from mechanics.

The charge and its conservation is the anchor of electrodynamics. Its
current $\mathcal{J}^k$ defines, by means of the Lorentz force
density $\mathfrak{f}_i$, the field strength $F_{ij}$, which allows to
define the magnetic flux $\Phi$. Faraday's induction law is an
incarnation of magnetic flux conservation.

Some people have no intuition about the conservation of a quantity
that is defined in 3 dimensions by integration over a 2-dimensional
area $\sim\int d\sigma^a B_a$, since we usually associate conservation
with a quantity won by 3-dimensional volume integration, namely
$\sim\int dV\rho$. Some mathematics education about dimensions will
enable us to understand the induction law as a ``continuity
equation.''

Summing up: the premetric Maxwell equations are a close-knit
structure, the 4-dimensional diffeomorphisms covariance holds it all
together. Clearly, a metric as well as a connection are alien to the
Maxwell equations.

\subsection{A local and linear electromagnetic response}

In order to fill the Maxwell equations with life, one has to relate
$F_{ij}$ to $\mathcal{H}^{ij}$:
\begin{equation}\label{Pconst0}
   \mathcal{H}^{ij}= \mathcal{H}^{ij}(F_{kl})
\end{equation}
If we assume this functional to be local, that is,
$\mathcal{H}^{ij}(x)$ depends only on $F_{kl}(x)$, and
linear homogeneously, then we find
\begin{equation}\label{constit1}
  \mathcal{H}^{ij}=\frac 12\,\chi^{ijkl}F_{kl}\qquad\text{with}\qquad 
  \chi^{ijkl}=-\chi^{ijlk}= -\chi^{jikl}\,;
\end{equation}
here the field $\chi^{ijkl}(x)$ represents the electromagnetic response
tensor density of rank 4 and weight $+1$, with the physical dimension
$[\chi]=1/\text{resistance}$. An antisymmetric pair of indices
corresponds, in 4 dimensions, to 6 independent components. Thus,
$\chi^{ijkl}$ can be understood as a $6\times 6$ matrix with 36
independent components.

We want to characterize the electromagnetic model spacetime by this
response tensor field $\chi^{ijkl}(x)$ with 36 independent
components.\footnote{Schuller et al. \cite{Schuller} took the
  $\chi^{ijkl}$--tensor density, which arises so naturally in
  electrodynamics, called the tensor proportional to it ``area
  metric'', and generalized it to $n$ dimensions and to string
  theory. For reconstructing a volume element, they have, depending on
  the circumstances, two different recipes, like, for example, taking
  the {\it sixth} root of a determinant. From the point of view of
  4-dimensional electrodynamics, the procedure of Schuller et al.\
  looks contrived to us.}  This is the tensor density defining the
{\it structure} of spacetime. It transcends the metric and/or the
connection.

We decompose the $6\times 6$ matrix into its 3 irreducible pieces. On
the level of $\chi^{ijkl}$, this induces
\cite{birkbook,Delphenich:2006bm}
\begin{eqnarray}\label{chidec}
  \chi^{ijkl}&=&\,^{(1)}\chi^{ijkl}+
  \,^{(2)}\chi^{ijkl}+
  \,^{(3)}\chi^{ijkl}\,.\\ \nonumber 36
  &=&\hspace{10pt} 20\hspace{10pt}\oplus \hspace{10pt}15\hspace{10pt}
  \oplus \hspace{11pt}1\,.
\end{eqnarray}
The third part, the {\it axion} part, is totally antisymmetric $^{(3)}
\chi^{ijkl}:= \chi^{[ijkl]} ={\a}\,{\epsilon}^{ijkl}$, with the
pseudoscalar ${\a}$, see also \cite{RaabSihvola}. The {\it skewon}
  part is defined according to $ ^{(2)}\chi^{ijkl}:=\frac
  12(\chi^{ijkl}- \chi^{kl ij})$.  Under reversible conditions,
  (\ref{constit1}) can be derived from a Lagrangian, then
  $^{(2)}\chi^{ijkl}=0$. The {\it principal} part $^{(1)}\chi^{ijkl}$
  fulfills the symmetries $ ^{(1)}\chi^{ijkl}= {}^{(1)}\chi^{klij}$
  and $^{(1)}\chi^{[ijkl]}=0$.

The local and linear response relation now reads
\begin{equation}\label{constit7}
  { \mathcal{H}^{ij}=\frac
    12\left({}^{(1)}{\chi}^{ijkl}+
      {}^{(2)}{\chi}^{ijkl} +{\a}\,
      {\epsilon}^{ijkl}\right)F_{kl}\,,}
\end{equation}
and, split in space and time \cite{birkbook,Hehl:2007ut},
\begin{eqnarray}\label{explicit'}
  {D}^a\!&=\!&( \varepsilon^{ab}\hspace{4pt} - \,
    \epsilon^{abc}\,n_c )E_b\,+(\hspace{7pt} \gamma^a{}_b +
    s_b{}^a - \delta_b^a s_c{}^c) {B}^b +
  {\alpha}\,B^a \,, \\ {H}_a\!&=\!&\>( \mu_{ab}^{-1}
    - {\epsilon}_{abc}m^c ) {B}^b +(- \gamma^b{}_a +
    s_a{}^b - \delta_a^b s_c{}^c)E_b -
  {\alpha}\,E_a\,;\label{explicit''}
\end{eqnarray}
here $\epsilon^{abc}= {\epsilon}_{abc}=\pm 1,0$ are the 3-dimensional
Levi-Civita symbols. The 6 permittivities $\varepsilon^{ab}=
\varepsilon^{ba}$, the 6 permeabilities $\mu_{ab}=\mu_{ba}$ were
already known to Maxwell. The 8 magnetoelectric pieces $\g^a{}_b$ (its
trace vanishes, $\g^c{}_c=0$) were found since 1961, see Astrov
\cite{Astrov}. Eventually, the hypothetical skewon piece
\cite{birkbook} carries 3 permittivities $n_a$, the 3 permeabilities
$m^a$, and the 9 magnetoelectric pieces $s_a{}^b$.  Equivalent
response relations were formulated by Serdyukov et al.\
\cite[p.86]{Serdyukov:2001} and studied in quite some detail, see also
de~Lange and Raab \cite{Lange:2015zja}.

Suppose we have as {\it special case} a vacuum spacetime described by
a Riemannian metric $g_{ij}$. Then the response tensor turns out to be
\begin{equation}\label{MaxLor}
  \chi^{ijkl}=\,^{(1)}\chi^{ijkl}=2\lambda_0\sqrt{-g}g^{i[k}g^{l]j}\qquad\text{and}
  \qquad \mathcal{H}^{ij}= \lambda_0\sqrt{-g}F^{ij}\,,
\end{equation}
with the vacuum admittance $\lambda_0\approx 1/377\, \Omega$. Thus, we
recover known structures, and we recognize that the relation
(\ref{constit1}) represents a natural generalization of the vacuum
case. The metric $g^{ij}$ can be considered as some kind of a square
root of the electromagnetic response tensor $\chi^{ijkl}$.

We should keep in mind that a local and homogeneous electromagnetic
response like (\ref{constit1}) can be, if the circumstances require
it, generalized to nonlocal and/or to nonlinear laws. Examples of {\it
  nonlocal} laws have been proposed by Bopp and Podolsky\footnote{See
  \cite[Sec.28-8]{Feynman}.} and by Mashhoon.\footnote{See
  \cite[Sec.E.2.2]{birkbook}.} {\it Nonlinear} laws are due to
Heisenberg and Euler,\footnote{See \cite[Sec.E.2.3]{birkbook}.} Born
and Infeld,\footnote{See \cite[Sec.E.2.4]{birkbook}.} and
Pleba\'nski.\footnote{See \cite[Sec.E.2.5]{birkbook}.} Fresnel
surfaces for the nonlinear case were found by Obukhov and Rubilar
\cite{Obukhov:2002xa}, for example. More recently, L\"ammer\-zahl et
al.\ \cite{Lammerzahl:2012kw} and Itin et al.\ \cite{Itin:2014uia}
investigated electrodynamics in Finsler spacetimes. In the premetric
framework, this corresponds to a nonlocal constitutive law, see
\cite[Eq.(3.29)]{Itin:2014uia}, somewhat reminiscent of the
Bopp-Podolsky scheme.

\subsection{Propagation of electromagnetic disturbances}

The obvious next step in evaluating the physics of our model of
spacetime is to look how electromagnetic disturbances propagate in
this spacetime. One can either consider the short wave-length limit of
the electromagnetic theory, the WKB-approximation, or one can study,
as we will do here, the propagation of electromagnetic disturbances
with a technique developed by Hadamard; for a general outline, see
\cite[Chap.~C]{Truesdell}.

Hadamard describes an elementary wave as a process that forms a wave
surface. Across this surface, the electromagnetic field is continuous,
but the derivative of the field has a jump. The direction of a jump is
given by the wave covector. The subsequent integration produces the
rays, with the wave vectors as tangents to rays, see for our case
\cite{Kiehn:1991,Fukui,birkbook,Itin:2009aa}. In the meantime, our
methods have been improved, see
\cite{Favaro,Baekler:2014kha,Favaro:2015jxa}.

Out of the electromagnetic response tensor density we can define, with
the help of the covariant Levi-Civita symbol $\epsilon_{ijkl}=\pm
1,0$, the premetric ``diamond'' (single) dual and the diamond double
dual, respectively:
\begin{equation}\label{doublediamond}
  \chi^{\diamond\, ij}{}_{kl}:=\frac 12 \chi^{ijcd}\epsilon_{cdkl}\,,\qquad
  {^\diamond}\chi^{\,\diamond}_{\hspace{2pt}ijkl}:=\frac 12  \epsilon_{ijab}
  \chi^{\diamond\, ab}{}_{kl} =\frac 14 \epsilon_{ijab}\chi^{abcd}
  \epsilon_{cdkl}\,.
\end{equation}
The covariant Levi-Civita symbol carries weight $-1$ and $\chi^{abcd}$
weight $+1$. Thus, the double dual has weight $+1$, too. Performing
the double dual apparently corresponds to a lowering of all four
indices of $\chi^{abcd}$---and this is achieved without having access
to a metric of spacetime.

After  this   preparation,  it   is  straightforward  to   define  the
(premetric) 4th  rank {\it Kummer  tensor density,} which is  cubic in
$\chi$, as \cite{Baekler:2014kha}
\begin{eqnarray}\label{KummerDef}
{{\cal K}^{ijkl}[\chi]}:= \chi^{ai bj}{}\,
{^\diamond}\chi^\diamond_{\hspace{2pt}acbd}\chi^{ckdl}\,.
\end{eqnarray}
It has weight $+1$ and obeys the symmetry ${{{\cal K}^{ijkl}}=\;{{\cal
    K}^{klij} \,.}}$

At each point in spacetime, the wave covectors $q_i=(\omega,\vec{k})$
of the electromagnetic waves span the Fresnel wave surfaces, which
are quartic in the wave covectors according to 
\begin{eqnarray}\label{Fresnel}
  {\cal K}^{ijkl}[\chi]\,q_iq_jq_kq_l
={\cal K}^{(ijkl)}[\chi]\,q_iq_jq_kq_l =0\,.
\end{eqnarray}
The Tamm-Rubilar (TR) tensor density \cite{Rubilar:2007qm,birkbook},
with the conventional factor $1/6$, is defined by
\begin{eqnarray}\label{TR}
 {\cal G}^{ijkl}\,[\chi]:=\frac{1}{6}\,{\cal K}^{(ijkl)}[\chi]
=\frac 16  \chi^{a(i j|b}{}\,
{^\diamond}\chi^\diamond_{\hspace{2pt}acbd}\chi^{c|kl)d}\,.
\end{eqnarray}
It is totally symmetric and
carries 35 independent components. By straightforward algebra it can
be shown that the axion field drops out from the TR-tensor:
\begin{equation}
  {\cal G}^{ijkl}\,[\chi]= {\cal G}^{ijkl}\,[\,^{(1)}\!\chi+{}^{(2)}\!\chi]\,;
\end{equation}
see in this connection also \cite{Itin:2007cv} and the references
given there. The effect of the skewon piece on light propagation has
been studied in \cite{Obukhov:2004zz}. Ni \cite{Ni:1977zz} was the first
to understand that the axion field doesn't influence the light
propagation in the geometrical optics limit. Note that
$^{(1)}\!\chi+{}^{(2)}\!\chi$ has $20+15$ independent components,
exactly as $\cal G$---probably not by chance.

Accordingly, the totally symmetric TR-tensor ${\cal G}^{ijkl}[\chi]$,
with its 35 independent components, can, up to a factor, be observed
by optical means, that is, the TR-tensor---in contrast to the Kummer
tensor, as far as we know---has a direct operational interpretation.

\subsection{Fresnel wave surface}

The (generalized) Fresnel equation
\begin{equation}\label{Fresnelx} {\cal
    G}^{ijkl}[\chi]\,q_iq_jq_kq_l=0\,,
\end{equation}
determines a Fresnel wave surface. A trivial test for checking the
correctness of ({\ref{Fresnelx}}) is to substitute the response tensor
for the Maxwell-Lorentz vacuum electrodynamics (\ref{MaxLor})$_{1}$
into the TR-tensor of ({\ref{Fresnelx}}). One finds straightforwardly
$(g^{ij}q_iq_j)^2=0$, that is, two light cones that collapse onto each
other. The decomposition of (\ref{Fresnelx}) into space and time can
be found in \cite[(D.2.44)]{birkbook}.

For illustration, following \cite{Schaefer,Baekler:2014kha}, see
also \cite{Itin:2009fj}, we will display a classical example of such a
surface.  In Eqs.(\ref{explicit'}) and (\ref{explicit''}), we choose
an anisotropic permittivity tensor with three different principal
values and assume trivial vacuum permeability, whereas all
magnetoelectric moduli---with the possible exception of the axion
$\alpha$---vanish,
\begin{equation}\label{figure1}
  (\varepsilon^{ab})=\left(\begin{matrix}\varepsilon^{1}&0&0\\0&\ve^{2}&0\\
      0&0&\ve^{3}\end{matrix}\right)\quad\text{and}\quad
  (\mu^{-1}_{ab})=\mu^{-1}_0\left(
    \begin{matrix}\hspace{2pt}1\hspace{2pt}\;&0&0
      \\0&\hspace{2pt}1\hspace{2pt}\;&0\\
      0&0&\hspace{2pt}1\hspace{2pt}\;\end{matrix}\right).
\end{equation} 
Substitution into the Fresnel equation yield the quartic polynomial
\begin{eqnarray}\label{FresnelQuadric}
&&\hspace{-20pt}(\a^2x^2+\b^2y^2+\g^2z^2)(x^2+y^2+z^2) \nonumber\\
&&\hspace{-18pt}-\left[\a^2(\b^2+\g^2)x^2+\b^2(\g^2+\a^2)y^2+\g^2(\a^2+\b^2)z^2
 \right]+\a^2\b^2\g^2=0\,,
\end{eqnarray}
with the 3 parameters\footnote{Here, in this context, $\alpha$ is {\it
    not} the axion field!}
$\a:=c/\sqrt{\ve_{1}},\;\b:=c/\sqrt{\ve_{2}},\;\g:=c/\sqrt{\ve_{3}}$,
and with $c=1/\sqrt{\ve_0\mu_0}$ as the vacuum speed of light.

The corresponding surface is drawn in Fig.1. As an example of a
Fresnel surface for a more exotic material, we provide one for the
so-called PQ-medium of Lindell \cite{Lindell:PQ}. It may turn out that this
response tensor can only be realized with the help of a suitable
metamaterial, see \cite{Shamonina}. Corresponding investigations are
underway by Favaro \cite{Favaro:2015privcomm}.

Let us shortly look back on what we have achieved so far: We have
formulated the Maxwell equation in a premetric way. For the response
tensor only local and linear notions are used, no distances or angles
were mentioned nor implemented. Under such circumstances,
electromagnetic disturbances propagate in a birefringent way in
accordance with the Fresnel wave surfaces, such as presented in Figs.1
and 2.

How can we now bring in distances and angles, which are concepts
omnipresent in everyday life? The answer is obvious, we have to
suppress birefringence.

\begin{figure}
\includegraphics[scale=.65]{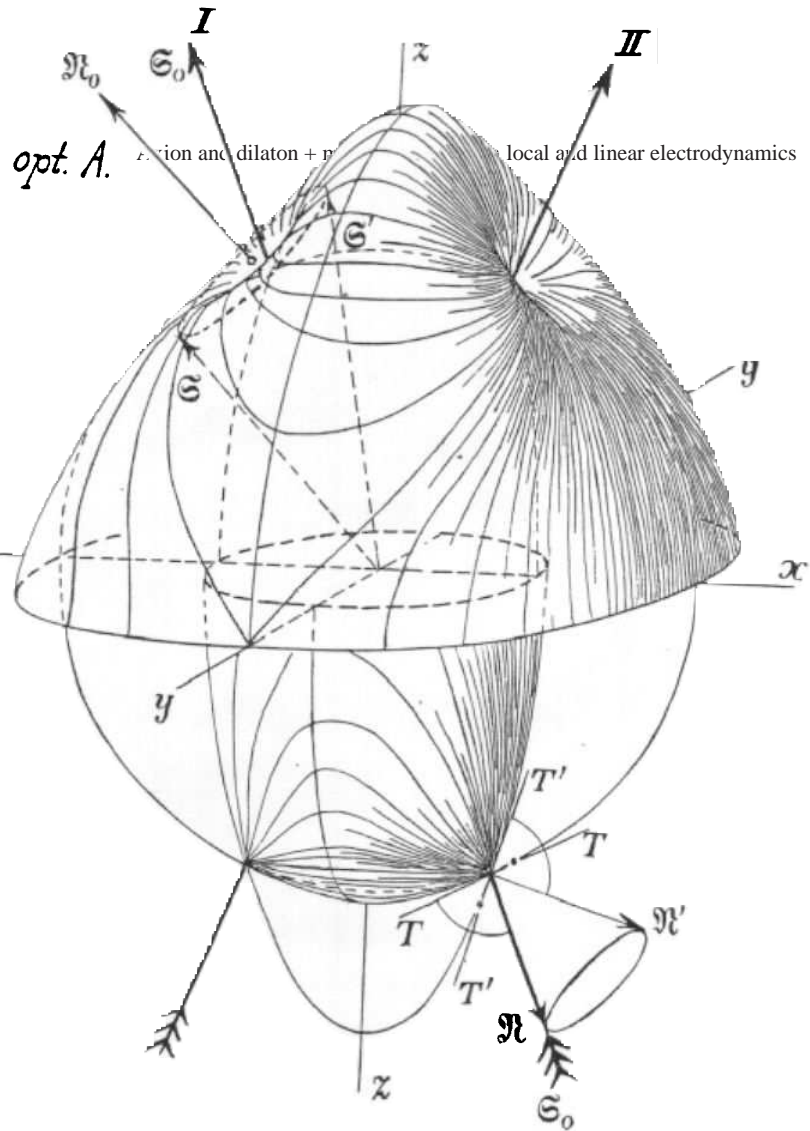}\\

\caption{Fresnel wave surface for the permittivities and the
  permeabilities of Eq.(\ref{figure1}). {\it It had been drawn by
    Jaumann for an optically biaxial crystal, see Schaefer
    \cite[p.485]{Schaefer}. This crystal has the property of
    birefringence (or double refraction). The origin at $x=y=z=0$ is
    the point in 3-dimensional space from where the wave covectors
    $\vec{k}$ originate. They end on the Fresnel wave surface. Their
    modulus is proportional to the reciprocal of the phase velocity
    $\omega/k$. In other words, up to a sign, we have usually in one
    direction two different phase velocities. This is an expression of
    the birefringence. Only along the optical axes {\bf I} and {\bf
      II}, we have only one wave covector. The upper half depicts the
    exterior shell with the funnel shaped singularities, the lower
    half the inner shell. The two shells cross each other at four
    points forming cusps.}}
\end{figure}
\begin{figure} 
 \includegraphics[width=0.8\textwidth=1.0]{
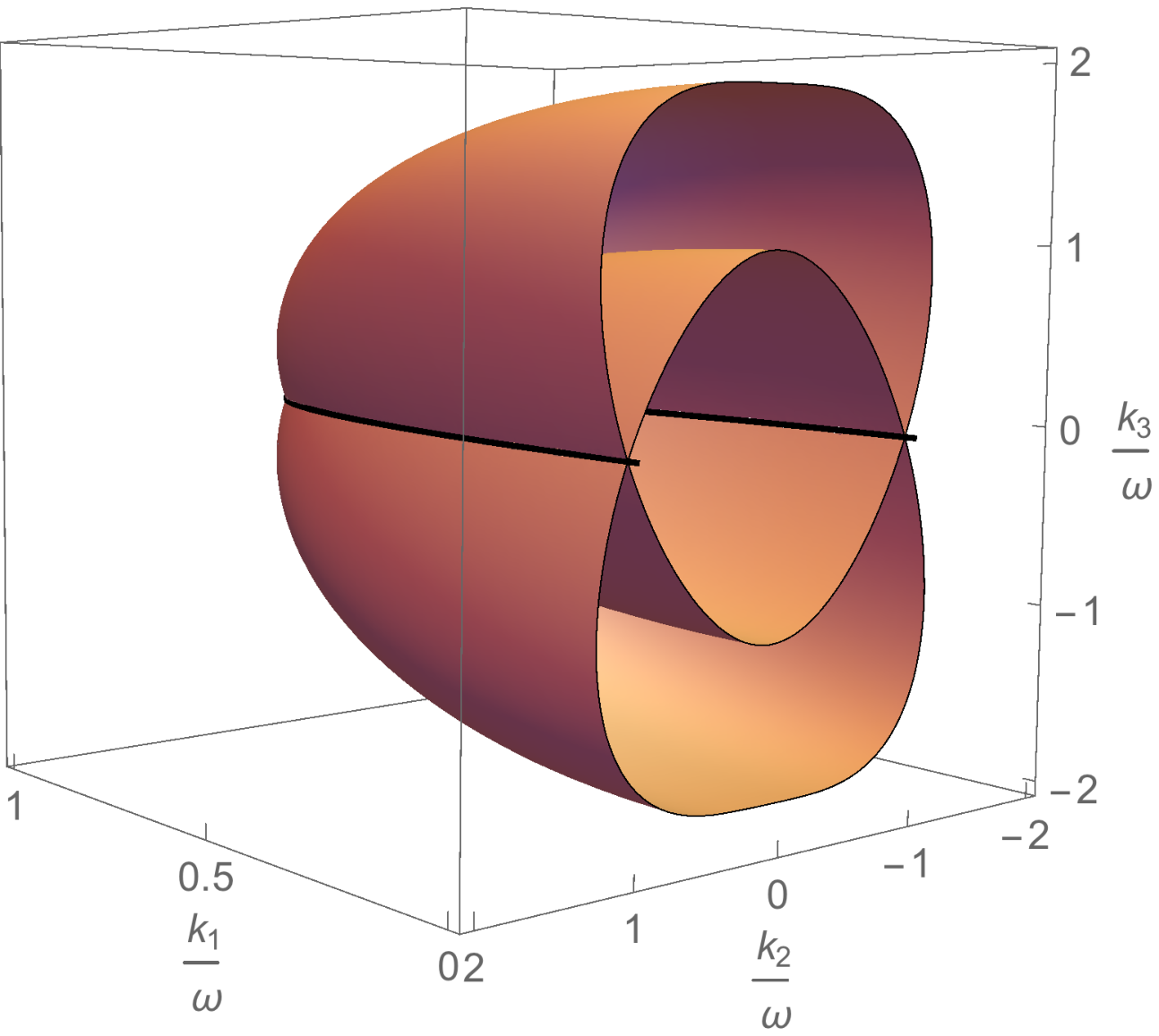}
\caption{Fresnel wave surface for a PQ-medium of Lindell
  \cite{Lindell:PQ,LindellFavaro}. {\it Using Lindell's dyadic version
    of the Fresnel equation \cite{Lindell:2005,Lindell:2015}, Sihvola
    \cite{Sihvola:2015privcomm} drew the Fresnel wave surface by using
    Mathematica. Our image was later created by Favaro
    \cite{Favaro:2015privcomm} in a similar way, again with
    Mathematica. For the wave covector, we have
    $q_i=(\omega,k_1,k_2,k_3)$.}}
\end{figure}

\subsection{Suppression of birefringence: the light cone}

Looking at the figures, it is clear that we have to take care that
both shells in each Fresnel wave surface become identical
spheres. Then light propagates like in vacuum. For this purpose, we
can solve the quartic Fresnel equation (\ref{Fresnelx}) with respect
to the frequency $q_0$, keeping the 3--covector $q_a$ fixed. One finds
four solutions, for the details please compare
\cite{Lammerzahl:2004ww,Itin:2005iv}. To suppress birefringence, one
has to demand {\it two conditions.} In turn, the quartic equation
splits into a product of two quadratic equations proportional to each
other. Thus, we find a light cone $g^{ij}(x)\,q_iq_j=0$ at each point
of spacetime.

Perhaps surprisingly, we derived also the {\it Lorentz signature,} see
\cite{birkbook,Itin:2004qr,ItinFriedman}.  This can be traced back to
the Lenz rule, which determines the relative sign of the two terms in
the induction law, as compared to the relative sign in the
Amp\`ere-Maxwell law. The Lorentz signature can be understood on the
level of classical electrodynamics, no appeal to quantum field theory,
which is widespread in the literature, is necessary.

Globally in the cosmos, birefringence is excluded with high accuracy,
see the observations of Polarbear \cite{Ade:2015cao} and the
discussion of Ni \cite{Ni:2014qfa}.

\subsection{Axion, dilaton, metric}

At the premetric level of our framework, besides the principal piece,
first the skewon and the axion fields emerged. Only subsequently the
light cone was brought up. The skewon field was phased out by our
insistence of the vanishing birefringence in the vacuum. Accordingly,
the axion field and the light cone survived the suppression of the
birefringence.

The light cone does not define the metric uniquely. Rather an
arbitrary function $\lambda(x)$ is left over:
\begin{equation}\label{conf}
  {\lambda}(x)\,g^{ij}(x)\,q_iq_j=0\,.
\end{equation} 
The light cone is invariant under the 15-parametric conformal
group. The 4 proper conformal transformation correspond to a
reflection at the unit circle and, as such, are of a nonlocal nature. As
a consequence, if two frames are related to each other by a proper
conformal transformation and one frame is inertial, the other one is
accelerated with respect to the former one. Accordingly, there is an
operational distinction possible between a proper conformal and a
dilation or scale transformation.  Thus, only the 11 parameter Weyl
subgroup of the 15 parameter conformal group is based on {\it local}
transformations.

If we compare our result in (\ref{conf}) with vacuum response in
(\ref{MaxLor}), we recognize, not forgetting the axion field, that we
find the following response equation for vanishing birefringence:
\begin{equation}\label{response}
  \mathcal{H}^{ij}=\boldsymbol{[}\underbrace{\lambda(x)}_{\hbox{dilaton}}
  \!\!\sqrt{-g}\,g^{ik}(x)\,g^{jl}(x)+\underbrace{\alpha(x)
    }_{\hbox{axion}} \epsilon^{ijkl}\,\boldsymbol{]} \,F_{kl}\,.
\end{equation}
Because of the presence of the dilation within the Weyl group, it is
natural to identify the function $\lambda(x)$ with the dilaton
field.\footnote{In the early 1980s, Ni \cite{Ni:1984} has shown the
  following: Suppressing the birefringence is a necessary and
  sufficient condition for a Lagrangian based constitutive tensor to
  be decomposable into metric+dilaton+axion in a weak gravitational
  field (weak violation of the Einstein equivalence principle), a
  remarkable result. Note that Ni assumed the existence of a
  metric. We, in (\ref{response}), derived the metric from the
  electromagnetic response tensor density $\chi^{ijkl}$.}

In the calculus of exterior differential forms, see \cite{birkbook},
the twisted excitation 2-form $H=\frac 12 \epsilon_{ijkl}{\mathcal
  H}^{kl}dx^i\wedge dx^j$ and the untwisted field strength 2-form
$F=\frac 12 F_{ij} dx^i\wedge dx^j$, together with the twisted current
3-form $J=\frac{1}{3!}\epsilon_{ijkl}\mathcal{J}^ldx^i\wedge dx^j
\wedge dx^k$, obey the Maxwell equations $dH=J$ and $dF=0$. By means
of the metric, we can introduce the Hodge star ${}^\star$
operator. Then the response relation (\ref{response}) becomes even
more compact \cite{Hehl:2005hu,Favaro:2010ys}:
\begin{equation}\label{response*}
  {H}=\boldsymbol{[}\lambda(x)\,^{\boldsymbol{\star}} +
  \alpha(x)\boldsymbol{]}F\,.
\end{equation}

Eqs.(\ref{response}) and (\ref{response*}) represent the end result of
investigating an electromagnetic spacetime model with local and linear
response and without birefringence. The three fields $\lambda(x)$,
$g^{ij}(x)$, and $\alpha(x)$ come up together with a reasonable
interpretation. At least in the way we defined them here,
$\lambda(x)$, $g^{ij}(x)$, and $\alpha(x)$ are all three {\it
  descendants of electromagnetism.}

As we have argued in Sec.1.5, the dilaton seems to be at home in the
Weyl-Cartan spacetime. Our results (\ref{response}) or
(\ref{response*}) are consistent with this expectation, that is, we
believe that these equations are valid in a Weyl-Cartan spacetime.

What are we told by experiments and observations? The axion $A^0$ has not
been found so far, so we can provisionally put $\alpha=0$.  Moreover,
under normal circumstances, the dilaton seems to be a constant field
and thereby sets a certain scale, that is, $\lambda(x)=
\lambda_{0}=\text{const}$, where $\lambda_{0}$ is the admittance of
free space, the value of which is, in SI-units, $\approx\!\!
1/(377\;\Omega)$. Under these conditions, we are left with the
response relation of conventional Maxwell-Lorentz electrodynamics,
\begin{equation}
  \mathcal{H}^{ij}
  =  \lambda_0\sqrt{-g}\,F^{ij}\qquad \hbox{or}\qquad  H=\lambda_0\,^\star F\,.
\end{equation}
The possible generalizations are apparent.

\section{Discussion}

Gravity, coupling to all objects carrying energy-momentum, is a truly
universal interaction. Electromagnetism is only involved in
electrically charged matter. What is curious and what we still do not
understand is that the gravitational potential $g^{ij}$ emerges in an
electromagnetic context, that is, in studying electromagnetic
disturbances, we can suppress birefringence, and then the light cone
emerges. And the light cone is essentially involved in general
relativity. In other words, we cannot formulate a general-relativistic
theory of gravity unless some electric charge is around:
electromagnetic waves are a necessary tool for constructing general
relativity.

Perlick is not concerned about it. He observes that
\cite{Perlick:2015} ``...the vacuum Maxwell equations are but one
example that have the light cones of the spacetime metric for their
characteristics.  The same is true of the Dirac equation, the
Klein-Gordon equation and others....'' Yes, this is true. However, if
a metric is {\it not} prescribed, we cannot even formulate Dirac's
theory. In contrast, in premetric electrodynamics, if a local and
linear response tensor density is assumed, we can derive the metric,
as we discussed above. In this sense, electrodynamics is distinguished
from Dirac's theory---and in this, and only in this sense, the
premetric Maxwell equations are more fundamental than the Dirac
equation.

Accordingly, there seems to be a deep connection between
electromagnetism and gravity, even though gravity is truly universal,
in contrast to electrodynamics.

\begin{acknowledgement}
  This project was partly supported by the German-Israeli Research
  Foundation (GIF) by the grant GIF/No.1078-107.14/2009. I am grateful
  to Claus Kiefer (Cologne) for helpful remarks re Higgs cosmology, to
  Yuri Obukhov (Moscow) re questions of torsion and of the Hadamard
  method, to Helmut Rumpf (Vienna) re the Mukhanov bound of
  $10^{-29}\,m$, and to Volker Perlick (Bremen) re gravity versus
  electromagnetism and what is more fundamental. I would like to thank
  Ari Sihvola and Ismo Lindell (both Espoo/Helsinki) and Alberto
  Favaro (London) for the permission to use the image in Fig.2. Many
  useful comments on a draft of this paper were supplied by Hubert
  Goenner (G\"ottingen), Yakov Itin (Jerusalem), Claus L\"ammerzahl
  (Bremen), Ecardo Mielke (Mexico City), Wei-Tou Ni (Hsinchu), Erhard
  Scholz (Wuppertal), Frederic Schuller (Erlangen), and Dirk Puetzfeld
  (Bremen). I am grateful to Alan Kosteleck\'y (Bloomington) for a
  last-minute exchange of emails.
\end{acknowledgement}



\end{document}